\documentclass[aps,twocolumn,secnumarabic,balancelastpage,amsmath,amssymb,nofootinbib,floatfix,10pt,pra]{revtex4-2}

\usepackage{graphicx}      
\usepackage{bm}            

\usepackage{booktabs}
\usepackage{siunitx}
\usepackage{placeins}
\usepackage{subfigure}
\usepackage{textcomp,gensymb}
\usepackage{braket}
\usepackage{xcolor}
\usepackage{comment}

\usepackage[colorlinks=true]{hyperref}  

\newcommand{\w}{\omega}
\newcommand{\W}{\Omega}

\newcommand{\bb}{\langle}
\newcommand{\rr}{\rangle}
\newcommand{\g}{\gamma}

\newcommand{\p}{\partial}
\newcommand{\mb}{\mathbb}
\newcommand{\al}{\alpha}
\newcommand{\bs}{\mathbf}
\newcommand{\m}{\mathcal}

\newcommand{\ep}{\varepsilon}
\newcommand{\eps}{\epsilon}

\newcommand{\n}{\nabla}

\begin{document}
\title{A Contactless Heat Engine Driven by Nonreciprocal Fluctuation-Induced Torques}

\author{Dhruv Shah}
\email[Correspondence email address: ]{dhruvashah2004@gmail.com}
\affiliation{Department of Physics, Massachusetts Institute of Technology, Cambridge, MA 02139, USA}
\author{Kiryl Asheichyk}
\email{kiryl.asheichyk@gmail.com}
\affiliation{Department of Theoretical Physics and Astrophysics, Belarusian State University, 5 Babruiskaya Street, 220006 Minsk, Belarus}
\author{David Gelbwaser-Klimovsky}
\email{dgelbi@technion.ac.il}
\affiliation{Schulich Faculty of Chemistry and Helen Diller Quantum Center, Technion--Israel Institute of Technology, Haifa 3200003, Israel}
\author{Noah Graham}
\email{ngraham@middlebury.edu}
\affiliation{Department of Physics, Middlebury College, Middlebury, VT 05753, USA}
\author{Mehran Kardar}
\email{kardar@mit.edu}
\affiliation{Department~of Physics, Massachusetts Institute of Technology, Cambridge, MA 02139, USA}
\author{Matthias Kr\"uger}
\email{matthias.kruger@uni-goettingen.de}
\affiliation{Institute for Theoretical Physics, Georg-August-Universit\"at G\"ottingen, 37077 G\"ottingen, Germany}
\date{\today}

\begin{abstract}
We describe a contactless heat engine in which quantum and thermal electromagnetic fluctuations act as the working medium. The setup consists of two concentric cylinders held at different temperatures. The inner cylinder stably levitates within the outer one due to repulsive nonequilibrium Casimir forces. The chirality of the setup is broken by using nonreciprocal dielectric materials, akin to application of a magnetic field along the common cylinder axis. Using Rytov fluctuational electrodynamics, we show that heat transfer and torque can be expressed in terms of an angular-momentum-resolved heat flux density, $\Phi_n(\omega)$: each exchanged photon carries energy $\hbar\omega$ and angular momentum $\hbar n$. In reciprocal media  contributions from modes $n$ and $-n$ cancel and there is no net torque; nonreciprocity breaks this symmetry and powers rotation of the inner cylinder.  Even in the absence of contact, electromagnetic fluctuations produce a frictional torque opposing rotation that we compute. This enables computation of characteristic steady state rotations, and estimation of the engine efficiency (which remains bounded by the Carnot limit). The cylindrical setup provides a natural realization of fluctuation-induced angular-momentum transfer and  a possible route toward nanoscale contactless engines.
\end{abstract}

\maketitle
\subsection*{Introduction}
Fluctuation-induced forces provide a natural connection between quantum field theory, statistical mechanics, and soft matter, and were a central theme in the work of Rudi Podgornik. On the quantum side, Podgornik and collaborators extended Lifshitz--van der Waals theory to multilayered, stratified, and inhomogeneous dielectrics, and later to fluctuation-induced torques between anisotropic layered materials~\cite{PodgornikHansenParsegian2003,PodgornikParsegian2004,LuPodgornik2016}. In the  soft-matter domain, he showed how thermal fluctuations of order parameters, charges, and hydrodynamic fields generate classical pseudo-Casimir forces  in confined liquid crystals, charged dielectrics, and fluids~\cite{ZiherlPodgornikZumer1999,ZiherlHaddadanPodgornikZumer2000,NajiDeanSarabadaniHorganPodgornik2010,MonahanNajiHorganLuPodgornik2016}. Here, motivated by this broad view of fluctuation-induced phenomena, we consider how nonequilibrium electromagnetic fluctuations, when combined with nonreciprocal material response, can produce torque and perform work.

The canonical quantum example of a fluctuation-induced force is the attractive Casimir force between mirrors separated by vacuum~\cite{cas}. 
It was generalized by Lifshitz theory to material media and finite temperatures~\cite{liftz}, and describes how resonances in the dielectric function encode the crossover to the near-field van der Waals regime~\cite{PodgornikParsegian2004}. Away from thermal equilibrium, the theoretical treatment of electromagnetic fluctuations by the Rytov formalism~\cite{rytov1959theory}, describes phenomena such as near-field radiative heat transfer~\cite{Polder_HT_plates}. 
Another surprising effect is the friction of vacuum, which for slow motions can be computed by Green-Kubo relations  within linear response~\cite{linearresponse,greenkubo}.
For a collection of dielectric objects held at different temperatures, the
pairwise radiative heat flux can be computed within the scattering formulation
of fluctuational QED~\cite{Messina2011,trace, Rodriguez_numericalArbitrary}.  When the objects are made of
nonreciprocal media~\cite{fanfan}, nonequilibrium electromagnetic fluctuations can also generate propulsive forces \cite{gyrodparticle}.  In this way, thermal radiation can transfer not only energy but also directed linear or angular momentum.

These ingredients were used in  Ref.~\cite{nr} to propose a heat engine based on two nonreciprocal parallel plates, with the vacuum EM field as a working medium. At unequal temperatures, the plates experience a
propulsive tangential force, allowing heat flow to be converted into mechanical
work through relative sliding motion. 
Here we develop the corresponding engine in a cylindrical geometry.  The setup consists of two concentric cylinders whose dielectric tensors are nonreciprocal but preserve azimuthal symmetry.  The temperature difference now drives both a heat flux and a propulsive torque;
when the outer cylinder is held fixed, this torque can rotate the inner cylinder
and  generate work.

We first introduce the cylindrical setup and the gyrotropic dielectric response
used to break reciprocity while preserving the angular-momentum quantum number. 
We then use the Rytov formalism to calculate the heat flux and propulsive torque between the cylinders in terms of an angular-momentum-resolved flux density
$\Phi_n(\omega)$~\cite{nr}. This quantity can be interpreted as 
spectral channel flux: a mode labelled by frequency $\omega$ and azimuthal index
$n$ carries energy $\hbar\omega$ and angular momentum $n\hbar$.  This common
mode-resolved structure immediately yields a bound relating torque and heat
flux.  We compare the resulting formulas to those for nonreciprocal parallel
plates, making explicit the correspondence between translational momentum in
the plate geometry and angular momentum in the cylindrical geometry.

We then analyze the symmetry requirements for propulsion.  In an azimuthally
symmetric reciprocal system, the contributions of the $n$ and $-n$ channels are
equal, so the net torque cancels.  Nonreciprocity breaks this symmetry and
allows the exchanged radiation to carry a net angular momentum.  We also
consider the radial stability of the concentric configuration.  Using the
Proximity Force Approximation (PFA), we argue that the inner cylinder is stable
when the corresponding parallel-plate interaction is repulsive, and we provide
analytic expressions for this nonequilibrium repulsion, extending results for
reciprocal plates~\cite{Bimonte2011}.  Finally, we include the contactless frictional torque on the rotating cylinder.
We compute this friction and compare it with the result from linear response
theory~\cite{linearresponse}. From these results, we obtain the engine efficiency and demonstrate that it is Carnot bounded.

\begin{figure}
    \centering
    \includegraphics[width=0.95\linewidth]{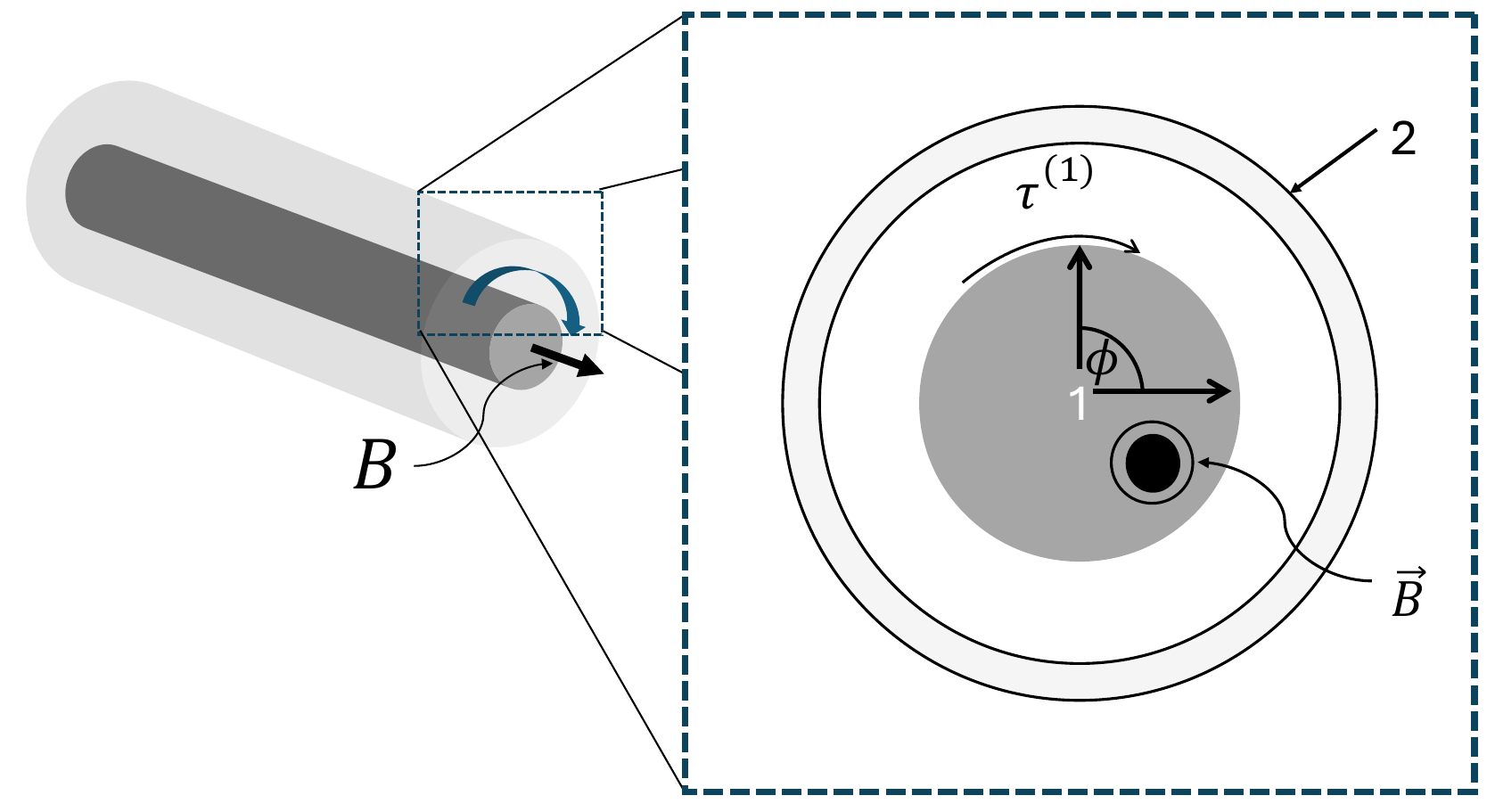}
    \caption{Two concentric cylinders, made of nonreciprocal dielectrics. The nonreciprocity is encoded through an external magnetic $\vec B$ field along cylinder axis $\hat z$. We study the torque $\tau^{(1)}$ acting on the inner cylinder when the cylinders are at unequal temperatures.}
    \label{fig:setup}
\end{figure}

\subsection*{Cylindrical geometry and nonreciprocal material}

The following symmetry assumptions are the minimal ingredients needed to make
angular momentum a good quantum number.  The setup consists of
concentric cylinders of radii
$R_1 < R_2$ along the $z$-axis, held at temperatures $T_1<T_2$
(see Fig.~\ref{fig:setup}) and consisting of nonmagnetic, nonreciprocal,
stationary, spatially local dielectrics.
(We use cylindrical coordinates $\vec r = (\rho,\phi,z)$.) 
Nonreciprocity can be due to internal symmetry breaking of the material, or induced by an external magnetic field $B_\text{ex}\hat z$; for simplicity we use $B_{\text{ex}}$ to encode the degree of nonreciprocity~\cite{caloz_electromagnetic_2018,
asadchy_tutorial_2020}. (Note that the nonreciprocity axis of $\hat z$ preserves azimuthal symmetry.) The two cylinders may have different dielectric
parameters, but in both cases the local dielectric tensor for cylinder
$i \in \{1,2\}$ has the form
\begin{align}\label{eq:dielectric tensor}
    \eps_{\text{nr},i}(\w)
    = \eps_{d,i} \m I
    + (\eps_{p,i}-\eps_{d,i})\hat z \otimes \hat z
    + i\eps_{f,i} \hat z\times \, .
\end{align}
Here $\eps_{p,i}$, $\eps_{d,i}$, and $\eps_{f,i}$ are complex functions of
$\w$ and of the cyclotron frequency $\w_c \propto B_\text{ex}$. In particular,
$\eps_{f,i}$ is the nonreciprocal parameter and is odd in $B_\text{ex}$.  A
specific example of these parameters for doped InSb is given in
Ref.~\cite{zhu_persistent_2016}.

Because the dielectric tensor $\eps_\text{nr}(\w)$ is azimuthally symmetric and
translationally invariant along $z$, electromagnetic operators derived from
$\eps_\text{nr}(\w)$ are block diagonal in the azimuthal eigenvalue
$n\in \mathbb{Z}$ and the longitudinal wavevector $k_z$.

\subsection*{Fluctuating currents and field correlators}
We next summarize only the parts of fluctuational QED needed to express heat and angular momentum transfer.
In the finite-temperature Rytov formalism, a dielectric contains fluctuating current sources. In equilibrium, the current correlators are related to the dielectric tensor $\ep(\w)$ by the fluctuation-dissipation theorem~\cite{eckhardt}. Using Maxwell's equations, one can relate the electric field $\bs E(\w)$ to the current sources, which yields a nonzero equilibrium spectral density for the symmetrized electric field correlations,
\begin{align}\label{eq: EE correlator}
    \bb \bs E \otimes \bs E^* \rr^\text{eq}_\w = a_T(\w) \, \frac{\mb G(\w)-\mb G^\dagger(\w)}{2i}\,,
\end{align}
where we have suppressed the spatial arguments and 
$\mb G(\w)$ is the electromagnetic Green's function for the system.  Also, $a_T(\w) \equiv 4\pi\hbar \w^2/c^2 \, \text{sgn}(\w) + 8\pi\hbar\w^2 \text{sgn}(\w)\nu_T(\w)/c^2$, wherein the first term represents zero-point QED fluctuations, while the second term represents finite-temperature contributions, given in terms of the standard Bose-Einstein density $\nu_T(\w)\equiv \left(e^{\hbar |\w|/k_{\textrm{B}}T}-1 \right)^{-1}$.
From Maxwell's equations, the Green's function satisfies the relation $\left[ \n\times \n \times -\w^2/c^2 \mb I - \mb V\right]\mb G = \mb I$, where $\mb V$ is the dielectric potential $\mb V(\omega) = ( \ep(\w)-1)\,\w^2/c^2 $. For our local dielectric, we write $\mb V = \mb V_1+\mb V_2$, where $\mb V_j$ is nonzero only within cylinder $j$. Since our system is not in thermal equilibrium, we must treat the current sources within the two objects separately. The total nonequilibrium electric field correlator can then be decomposed as~\cite{pretrace}
\begin{align}
    \bb\bs E\otimes \bs E^*\rr_{\w}^\text{neq} = \bb\bs E\otimes \bs E^*\rr_{\w,1}(T_1)+\bb\bs E\otimes \bs E^* \rr_{\w,2}(T_2)\,,
\end{align}
where $\bb\bs E\otimes \bs E^*\rr_{\w,j}$ includes fluctuations from within cylinder $j$. Because the outer cylinder is considered to be thick, the environment has no effect on the fields in the gap between the cylinders. The total Green's function for the system can then be expressed in terms of the electromagnetic scattering operators $\mb T_j(\w)$ of the two cylinders, which are defined as $\mb T_j = \mb V_j \left(1-\mb G_0\mb V_j \right)^{-1}$, where $\mb G_0$ is the vacuum Green's function~\cite{trace}.

\subsection*{Heat and angular momentum flux}
The central point is that the same mode-resolved flux controls both energy transfer and torque.
We denote the heat flux and $z$-component of torque due to fluctuations from object $i$, which are absorbed by or acting on object $j$, as $H_i^{(j)}$ and $\tau_i^{(j)}$, respectively. For simplicity, we initially consider $i=2$ and $j=1$, i.e.\ the heat and angular momentum flux from the outer to the inner cylinder. The heat flux can be computed by integrating the normal component of the Poynting vector on a surface $\Sigma_1$ enclosing the inner and excluding the outer cylinder. Similarly, the torque can be expressed as a surface integral in terms of the stress tensor $T_{ij}\equiv E_iE^*_j+B_iB^*_j-\delta_{ij}\left(|E|^2+|B|^2\right)/2$. For convenience, we pick $\Sigma_1$ to be an enclosing cylinder of radius $R_1+\delta$ with $\delta \rightarrow0^+$. We obtain
\begin{align}\label{eq:surface int}
    H_2^{(1)}&=\int_{-\infty}^\infty \frac{d\w}{2\pi} \int_{\Sigma_1}\frac{c^2\hat \rho}{4\pi \w} \cdot \text{Im}\bb \bs E\times \left(\n \times \bs E^* \right)\rr_{\w,2}  \\
    \tau_2^{(1)} &= \int_{-\infty}^\infty \frac{d\w}{2\pi}\int_{\Sigma_1} \frac{R_1\hat \rho}{4\pi} \cdot \text{Re}\bb \bs E \otimes \bs E^* +\bs B\otimes \bs B^* \rr_{\w,2}\cdot \hat \phi\,. \nonumber
\end{align}
Since both averages above are quadratic in the electric field, they are easily expressed in terms of the  correlator $\bb \bs E\otimes \bs E^*\rr_{\w,2}$, which in turn can be expressed in terms of $\mb G_0$ and the scattering operators $\mb T_1,$ and $\mb T_2$, as described above. By symmetry in $\phi$ and $z$, the integrands in Eq.~(\ref{eq:surface int}) are constant, so the correlator can be evaluated with both arguments at a single point $\vec r_1\in \Sigma_1$.

The scattering operators $\mb T_j$ are block diagonal in $n$ and $k_z$ in the basis of vector cylindrical wavefunctions  $\{\vec L_{n,k_z}(\vec r)$, $\vec M_{n,k_z}(\vec r)$, $\vec N_{n,k_z}(\vec r)\}$~\cite{wengchochew}. In vacuum, the electric field consists only of the transverse $M$ and $N$ polarizations. The scattering operators can thus be expressed in terms of the $2\times 2$ polarization space scattering matrices $\m T_{1,n,k_z}$ and $\m T_{2,n,k_z}$ for the two cylinders.

\begin{table*}[t]
    \caption{Heat flux, force, and equilibrium friction expressions in the parallel plate~\cite{nr} and cylinder geometry. Here $\nu_T'(\w) \equiv \p_\w \nu_T(\w)$. Note that the heat flux densities are extensive: $S_i^{(j)}\propto L^2$ and $\Phi_i^{(j)}\propto L$, where $L$ is length of the plate or cylinder.}
    \begin{ruledtabular}
        \label{hf table 2 geometries}
        \begin{tabular}{ccc}
             & $\text{Parallel plates}$ &  $\text{Concentric cylinders}$\\
             \hline
             Heat flux  & $H_2^{(1)}=-4\hbar\int_0^{\infty }\frac{d\w}{2\pi} \nu_{T_2}(\w) \w \int \frac{dk_y}{2\pi} S_{2}^{(1)}(k_y; \w)$ & $H_2^{(1)}=4\hbar\int_{0}^\infty\frac{d\w}{2\pi} \nu_{T_2}(\w) \w \sum_n \Phi_{2,n}^{(1)}(\w)$ \\
              Propulsive force/torque& $F_{y,2}^{(1)}=-4\hbar\int_0^\infty \frac{d\w}{2\pi} \nu_{T_2}(\w)  \int \frac{dk_y}{2\pi} k_yS_{2}^{(1)}(k_y;\w)$ & $\tau_2^{(1)}=4\hbar \int_{0}^\infty \frac{d\w}{2\pi}\nu_{T_2}(\w) \sum_n n\Phi_{2,n}^{(1)}(\w)$\\
              Equilibrium friction &
              $\g_{2}^{(1)} = 4\hbar \int_0^\infty\frac{d\w}{2\pi}\nu_{T_2}'(\w) \int\frac{dk_y}{2\pi}k_y^2 S_2^{(1)}(k_y,\w)$ &
              $\g_2^{(1)}=-4\hbar \int_0^\infty\frac{d\w}{2\pi} \nu_{T_2}'(\w)\sum_n n^2 \Phi_{2,n}^{(1)}(\w) $
        \end{tabular}
    \end{ruledtabular}
\end{table*}

\paragraph*{Angular-momentum-resolved flux:}
We find that the expectation values in Eq. ~(\ref{eq:surface int})  for heat flux and torque are given by
\begin{align}\label{eq: ht+torque in hfd}
H_2^{(1)}&=4\hbar\int_{0}^\infty\frac{d\w}{2\pi} \nu_{T_2}(\w) \w \sum_n \Phi_{2,n}^{(1)}(\w) \nonumber \\
    \tau_2^{(1)}&=4\hbar \int_{0}^\infty \frac{d\w}{2\pi}\nu_{T_2}(\w) \sum_n n\Phi_{2,n}^{(1)}(\w)\,.
\end{align}
These are written in terms of the `heat flux density per mode' $\Phi_{2,n}^{(1)}(\w)$, which can be thought of as the number of photons of energy $\hbar \w$ and angular momentum $n\hbar$ emitted by cylinder $2$ and absorbed by cylinder $1$. As a result, the energy and angular momentum flux are the product of the heat flux density, the Bose-Einstein factor, and $\hbar \w$ or $n\hbar$, respectively. Our calculation allows us to explicitly write $\Phi_{i,n}^{(j)}(\w)$ in terms of a trace over functions of the scattering matrices $\m T_{1,n,k_z}$ and $\m T_{2,n,k_z}$ for the cylinders~\cite{longerpaper}. 

It can be shown that $\Phi_{2,n}^{(1)}\geq0$~\cite{longerpaper}, so that $|\sum_{n=-N}^{N} n\Phi_{2,n}^{(1)}(\w)|\leq N \sum_{n=-N}^{N} \Phi_{2,n}^{(1)}(\w)$. This result gives a bound for the magnitude of the torque in terms of the heat flux, similar to the case of parallel plates~\cite{nr}, because the sum over $n$ can, in explicit computations, be truncated. For example, for a thin inner cylinder, only $n=0$ and $n=\pm 1$ contribute, and the bound holds with $N=1$.

\subsection*{From linear momentum to angular momentum}
The cylindrical result is the angular-momentum analogue of the earlier parallel-plate propulsion formula.
The formulas in Eq.~(\ref{eq: ht+torque in hfd}) can be compared with those for a pair of translationally symmetric plates oriented parallel to the $yz$-plane, as shown Table~\ref{hf table 2 geometries}. In the latter geometry, all EM operators are diagonal in the in-plane wave-vector $(k_y,k_z)$, and there is a corresponding heat flux density per mode $S_2^{(1)}(k_y)$. This heat flux density can be written in terms of the $2\times 2$ polarization space parallel plate scattering matrices~\cite{fanfan,nr}.

The heat flux and torque in Eq. ~\eqref{eq:surface int}  can also be expressed in terms of trace formulas~\cite{trace, Strekha22} over generic scattering operators. The corresponding properties for a more generic geometry and nonreciprocal dielectric will be analysed in a companion manuscript~\cite{longerpaper}. 

\paragraph*{Pairwise flux relations and detailed balance:}

Before discussing propulsion, it is useful to separate kinematic balance
relations from the symmetry breaking due to nonreciprocity. In our two-object
setup there are four heat flux densities $\Phi_{i,n}^{(j)}$, with
$i,j\in\{1,2\}$. By explicitly manipulating their expressions in terms of the
scattering matrices $\m T_i$, one finds
\begin{align}\label{eq:hfd symmetries}
    \Phi_{i,n}^{(1)}=-\Phi_{i,n}^{(2)} \,,  \qquad
    \Phi_{1,n}^{(j)}=-\Phi_{2,n}^{(j)}\,\, ; \, i,j\in \{1,2\}.
\end{align}
These relations have simple physical interpretations. The first follows from
the surface independence of the flux expression in Eq.~(\ref{eq:surface int}).
For a fixed source $i$, the heat flux onto the inner cylinder is obtained by
integrating the Poynting vector over an enclosing surface in the vacuum gap. The
same surface also gives the heat flux absorbed by the outer cylinder, with the
opposite sign because the surface normal $\hat n$ is reversed. The second
relation in Eq.~\eqref{eq:hfd symmetries} ensures that, in thermal equilibrium,
the inner cylinder is neither cooled nor heated.

The relations in Eq.~(\ref{eq:hfd symmetries}) allow us to write simple
expressions for the net heating rate
$H^{(1)} = H_1^{(1)}+H_2^{(1)}$ and net torque
$\tau^{(1)} = \tau_1^{(1)}+\tau_2^{(1)}$ acting on the inner cylinder:
\begin{align}
    H^{(1)}
    &=
    4 \int_0^\infty \frac{d\w}{2\pi}\sum_n
    \left[ \nu_{T_2}(\w)-\nu_{T_1}(\w)\right]
    \hbar  \w \Phi_{2,n}^{(1)}(\w) \nonumber\\
    \tau^{(1)}
    &=
    4 \int_0^\infty \frac{d\w}{2\pi}\sum_n
    \left[ \nu_{T_2}(\w)-\nu_{T_1}(\w)\right]
    \hbar  n \Phi_{2,n}^{(1)}(\w) \,.
\end{align}

\paragraph*{Why reciprocity forbids torque:}
When the cylinders are made of reciprocal media, i.e., $\epsilon_f=0$ in Eq.~\eqref{eq:dielectric tensor}, we have $\Phi_{i,n}^{(j)}(\w)=\Phi_{i,-n}^{(j)}(\w)$, due to the symmetry of the scattering matrix $\m T_{i,n,k_z}=\m T_{i,-n,k_z}$~\cite{trace}. As a consequence, the torque vanishes, $\tau^{(1)}=0$. Expanding for small $\eps_f$, we find that the effect of $\pm n$ symmetry breaking is linear in $\eps_f$, so $\tau^{(1)} \propto \eps_f$ for small $\eps_f$.

These results are expected for the present setup with the dielectric tensor in
Eq.~(\ref{eq:dielectric tensor}), since the geometry remains symmetric under
reversal of the direction of rotation. In the companion manuscript~\cite{longerpaper},
we show that even when this geometric symmetry is broken, the torque vanishes for reciprocal media.

\subsection*{Radial stability of the contactless configuration}
The inner cylinder is suspended so that it is concentric with the fixed outer cylinder. To study the stability of the free inner cylinder under small displacements from the $z$-axis, we use the Proximity Force Approximation (PFA)~\cite{pfa}. In the small separation limit $R_2-R_1\ll R_1$ the radial interactions between the cylinders can be approximated by those between a pair of closely spaced parallel plates. Given a power-law scaling with separation of the interaction energy per unit area of the plates, we can show that the inner cylinder is stable if the parallel plate interaction is repulsive~\cite{longerpaper}.

\paragraph*{Repulsive plate forces as a stability criterion:}
While the equilibrium pressure between reciprocal plates is generally attractive~\cite{equilbrium_attraction}, keeping dilute plates out of thermal equilibrium with temperatures $T_1\neq T_2$ offers the possibility for repulsion~\cite{Bimonte2011}. We investigate this case, but now with a nonreciprocal material described by the dielectric tensor in Eq.~\eqref{eq:dielectric tensor}. The total pressure on plate 1 can be decomposed as $P^{(1)}(T_1,T_2)=P_0^{(1)}+\sum_{i=1,2}P_i^{(1)}(T_i)$, where $P_0^{(1)}$ is the equilibrium zero-point Casimir attraction, and the $P_i^{(1)}$ are finite-temperature contributions from the two plates. General trace expressions for all of these pressures are known in terms of parallel-plate scattering operators or reflection matrices ~\cite{trace,nr,Zhao2009,Grushin2011,Iizuka2023}. We restrict to small separations $d\ll c/\w, c/\left(\w|\eps_{\textrm{nr},i}^{\alpha\beta}|^{1/2}\right)$ in the range of relevant frequencies $\w$. In this regime, the reflection matrices are dominated by the diagonal electric-electric component~\cite{nr}. 
We consider the Drude-Lorentz model for the dielectric response defining $\eps_{p,i}$, $\eps_{d,i}$, and $\eps_{f,i}$ for the two plates. These are functions of the material  parameters $\w_{i}$, $\w_{\textrm{p}}$, and $\g$:  the zero-field resonant frequency, plasma frequency, and damping rate, respectively \cite{Milton2023}. The resonance frequencies $\w_{i}\,,\, i\in\{1,2\}$ are assumed to be distinct for the two plates. We get the total pressure
\begin{align}
    P^{(1)} \simeq \frac{k_{\textrm{B}}T_2}{32\pi d^3}\frac{(\w_{\textrm{p}}/\w_2)^4}{\widetilde \w^2 (\widetilde \w+1)}\left[ \frac{-\hbar\w_1}{2k_{\textrm{B}}T_2}+\frac{\frac{T_1}{T_2}-\widetilde \w^2}{\widetilde \w -1}\right]+\m O(B_\text{ex}^2)\,,
\end{align}
where $\widetilde \w =\w_1/\w_2$ and we have assumed that $\w_2|\widetilde \w-1|\gg \g $ and $ \hbar\omega_i/(k_{\textrm{B}}T_i) \ll 1 $. The first term is the zero-point attraction, while the second term can be made repulsive if $1<\widetilde \w^2<T_1/T_2$ or $T_1/T_2<\widetilde \w^2<1$, and can overcome the zero-point attraction for small $|\widetilde \w-1|\gg \g/\w_2$. Nonreciprocity enters the pressure expression at order $B_\text{ex}^2$, and thus does not alter this discussion for small $B_\text{ex}$, as expected from symmetry: nonreciprocity affects the normal pressure at order $B_\text{ex}^2$, but the tangential force at order $B_\text{ex}$.

\subsection*{Slow rotation and contactless friction}
A functioning engine must rotate, and rotation inevitably produces a fluctuation-induced drag torque.
When the inner cylinder rotates at angular speed $\W$, we must consider dynamical effects on the heat transfer and torque. For small $\W$, both these quantities are modified at first order. We write $H^{(j)}(\W)= H^{(j)}(0)+\W h^{(j)}+\m O(\W^2)$ and $\tau^{(j)}(\W)=\tau^{(j)}(0)-\W\g^{(j)}+\m O(\W^2)$. Here $h^{(j)}, \g^{(j)}$ are the linear response coefficients, with the latter pertaining to the contactless friction. To calculate these quantities, we need the EE correlator in the case of a rotating inner cylinder and a fixed outer cylinder, which we obtain by Lorentz transforming  the inner cylinder operators from the comoving frame to the lab frame. For small $\W$, the relevant Lorentz transformation is $\phi=\phi'+\W t'$, $t=t'$, where the primed coordinates correspond to the comoving frame. This leads to corresponding transformations for the electric fields and dielectric tensor~\cite{movingcorr}.

Rotation does not break the azimuthal or $z$ translation symmetries, so the transformed effective potential $\mb V_1^\W$ and scattering operator $\mb T_1^\W$ are still diagonal in $\w$, $n$, and $k_z$. Their values in the $n$ subspace become functions of the shifted frequency $\w-n\W$~\cite{longerpaper}. Most importantly, knowing the rotating scattering operator $\mb T_1^\W$ allows us to use previously established expressions with the substitution $\mb T_1 \rightarrow \mb T_1^\W$ ~\cite{movingcorr, emissionrotating}. The rotating heat fluxes and torques can thus be expressed in terms of rotating heat flux densities $\Phi_{i,n}^{(j),\W}$, with the same trace expressions. The total torque on the inner cylinder is then
\begin{align}
    \tau^{(1)}(\W) = 4\hbar \int_0^\infty\frac{d\w}{2\pi}\sum_n\left[ \nu_{T_2}(\w)-\nu_{T_1}(\w-n\W)\right]n\Phi_{2,n}^{(1),\W}.
\end{align}
We see that the Bose-Einstein factor for the first cylinder also contains the shifted frequency. A similar equation holds for the heat transfer as well, which links $h^{(j)}$ to the expression for torque via the corresponding Onsager symmetry as described below. The above equation can be expanded to first order in $\W$ to extract the friction coefficient, which yields the sum of two terms, a thermal equilibrium term and a nonequilibrium term proportional to $\nu_{T_2}-\nu_{T_1}$. The equilibrium term is presented in Table \ref{hf table 2 geometries}, and is similar to that for parallel plates with the replacements $n \Leftrightarrow k_y$ and $\Phi_n \Leftrightarrow S(k_y)$. The equilibrium friction can also be calculated from linear response theory~\cite{linearresponse, thesisgolyk}. 

\subsection*{Power output and efficiency} 
We now have all the ingredients to compute the efficiency of the cylinder engine. To lowest nontrivial order in $\W$,
\begin{align}\label{eq:efficiency}
    \eta_\text{cyl} = \frac{\tau^{(1)}\W}{H^{(1)}} = \W\frac{\tau^{(1)}(0)-\W\g^{(1)}(0)}{H^{(1)}(0)+\W h^{(1)}(0)}\,.
\end{align}
It is easy to compute the optimal slow rotation speed $\W^*$ to maximize efficiency in Eq.\ (\ref{eq:efficiency}). For a small temperature difference $\Delta T\equiv |T_2-T_1|\ll T_1$, i.e., close to global thermal equilibrium, we can prove that $h^{(1)}\simeq (T_2/\Delta T)\, \tau^{(1)}(0)$~\cite{longerpaper}, which ensures that the Carnot bound $\eta_\text{cyl} \leq \Delta T/T_2$ is satisfied. This relation is expected from the Onsager symmetry relations between linear response coefficients in global thermal equilibrium~\cite{nr, onsager_recip1, onsager_recip2}. For an explicit estimate of the efficiency, we consider a material with $\Phi_{2,n}^{(1)}$  peaked at a single resonance frequency $\w^*$. We further take the thin inner cylinder limit, so that the dominant contributions to heat flux and torque come from the $n=0,\pm 1$ modes. We define the  dimensionless ratios $\al= \frac{\Phi_{n=1}-\Phi_{n=-1}}{\Phi_{n=0}+\Phi_{n=1}+\Phi_{n=-1}}$ and $\sigma = \frac{\Phi_{n=1}+\Phi_{n=-1}}{\Phi_{n=0}+\Phi_{n=1}+\Phi_{n=-1}}$,  with $\al^2<\sigma^2<1$. $\al$ is a measure of nonreciprocity, and it is proportional to $\eps_f$ for small $\eps_f$. 
At $\W^*  =\eta_c\w\sqrt{\frac{\sigma(x-1)}{x+1}}$, efficiency is maximal,  taking the value $\eta_\text{cyl}^*=\eta_c\frac{x-1}{x+1}\leq \eta_c$, with Carnot efficiency $\eta_c=\Delta T/T_2$ and $x=\sqrt{1+\al^2/\sigma}>1$. In the small nonreciprocity limit, the maximal efficiency is 
$\eta_\text{cyl}^* \simeq  \frac{\al^2}{4\sigma }\eta_c$.
Last, for $\Omega\ll\omega^*$, $\eta_\text{cyl}$ grows linearly in $\Omega$, and the bound between torque and heat flux implies a bound for the initial slope, i.e., $\partial_\Omega\eta_\text{cyl}|_{\Omega=0}\leq \frac{1}{\omega^*}$.

\section*{Conclusions and outlook}
We have developed a description of a contactless heat engine in which
finite-temperature QED fluctuations act as the working
medium. For a nonreciprocal inner cylinder enclosed by a fixed outer cylinder,
we have shown that a temperature difference produces both radiative heat
transfer and a propulsive torque. Both quantities can be expressed in terms of
an angular-momentum-resolved flux density, which describes modes carrying energy
$\hbar\w$ and angular momentum $n\hbar$. This formulation makes explicit the
correspondence between azimuthal symmetry in the cylindrical case and translational invariance in the parallel-plate geometry.

We have also analyzed the radial stability of the freely suspended inner cylinder. In the Proximity Force Approximation, stability requires repulsion between closely spaced parallel plates. We show that such repulsion can arise out of thermal equilibrium and that, for weak nonreciprocity, it is not strongly modified by the nonreciprocal response.
Finally, we examined the linear response to rotation of the inner cylinder and
computed the corresponding contactless friction coefficient. Combining this
friction with the propulsive torque and heat flux yields the efficiency of the
cylindrical engine, which is bounded by the Carnot limit and increases with the
degree of nonreciprocity.

Future work can investigate whether nonreciprocity is required for engine
operation in more general geometries, and how these fluctuation-induced torques
may be optimized in experimentally realizable materials.

\subsection*{Acknowledgements}
We would like to thank Daniele Gamba for useful discussions. This research project was financially supported by the state of Lower-Saxony and the Volkswagen Foundation, Hannover, Germany (M.Kr. and D.G.) and the National Science Foundation  through grants
PHY-2209582 (N.G.) and DMR-2218849 (M.Ka.).


\appendix

\bibliography{refs}

\end{document}